\documentclass[a4paper,11pt]{article}
\usepackage{pos}

\newcommand{\centeredgraphics}[2][]{\vcenter{\hbox{\includegraphics[#1]{#2}}}}

\title{Towards QCD at Five Loops}
\ShortTitle{Towards QCD at Five Loops}

\author*[a]{Andreas Maier}
\author[a]{Peter Marquard}
\author[b]{York Schröder}

\affiliation[a]{Deutsches Elektronen-Synchrotron DESY,\\
  Platanenallee 6, 15738 Zeuthen, Germany}

\affiliation[b]{
  Centro de Ciencias Exactas,
  Departamento de Ciencias Básicas,
  Universidad del Bío-Bío,
  Avenida Andrés Bello 720, Chillán, Chile
}

\emailAdd{andreas.martin.maier@desy.de}
\emailAdd{peter.marquard@desy.de}
\emailAdd{yschroder@ubiobio.cl}

\abstract{%
  We report on recent progress on five-loop calculations in
  perturbative QCD. We discuss the computation of the perturbative
  quark condensate, decoupling in QCD, and the precision determination
  of charm and bottom quark masses. For the latter, we give some first
  results from a gauge-invariant subset of five-loop diagrams.
}

\FullConference{Loops and Legs in Quantum Field Theory (LL2024)\\
 14-19, April, 2024\\
Wittenberg, Germany\\}

\begin{document}
\maketitle

\section{Introduction}

A cursory inspection of publications over the last two decades
suggests that the first five-loop QCD calculation was completed as
early as in 2005~\cite{Baikov:2005rw}, with more five-loop QCD results
coming out over the following
years~\cite{Schreck:2007um,Baikov:2008jh,Baikov:2010je,Baikov:2012er,Baikov:2012zn,Baikov:2014qja},
culminating in a flurry of activity in 2016 and
2017~\cite{Baikov:2016tgj,Luthe:2016xec,Herzog:2017ohr,Luthe:2017ttc,Baikov:2017ujl,Herzog:2017dtz,Luthe:2017ttg,Chetyrkin:2017bjc,Baikov:2018nzi},
and further work
afterwards~\cite{Herzog:2018kwj,Fael:2022frj}. However, a second look
reveals that almost all of these works use techniques like asymptotic
expansion~\cite{Chetyrkin:1988zz,Chetyrkin:1988cu,Smirnov:1990rz,Beneke:1997zp}
or the R*
operation~\cite{Chetyrkin:1982nn,Chetyrkin:1984xa,Smirnov:1985yck,Chetyrkin:2017ppe,Herzog:2017bjx}
to reduce the number of loops to at most four. The only exception is a
series of works~\cite{Luthe:2016xec,Luthe:2017ttc,Luthe:2017ttg}
introducing an unphysical auxiliary
mass~\cite{Misiak:1994zw,vanRitbergen:1997va,Chetyrkin:1997fm}, which
means that the theory is arguably no longer QCD. It can therefore be
argued that there are no genuine five-loop QCD results to date.

What could we hope to learn from a five-loop QCD calculation? The
works cited above give us a first idea about the type of quantities
that should be within reach soon. Order $\alpha_s^5$ corrections to
the Adler function and to total hadronic decay widths, e.g.\ of the
$\tau$ or of the Z boson, are in principle desirable for
determinations of the strong coupling. However, in general better data
will be needed for a significant increase in precision. Next, the
energy dependence of the strong coupling and the quark masses could be
determined to formally six-loop order. This would include decoupling
of heavy flavours at five loops. Another example is the calculation of
five-loop moments of the total cross section for heavy hadroproduction
at lepton colliders, which would greatly benefit charm- and bottom
quark mass determinations.

Aside from the general interest in the precise knowledge of
fundamental parameters, the masses of the charm and the bottom quark
have to be known with high precision for applications in flavour
physics and for future Higgs coupling measurements. Projections for
the HL-LHC suggest that the strength of the Yukawa coupling to the
bottom quark will be measured with statistical and systematic
experimental uncertainties of about one per
cent~\cite{Cepeda:2019klc}. At a high-energy lepton collider, the
overall uncertainty could be improved by at least a factor of two and
per cent level precision is within reach for the charm Yukawa
coupling~\cite{Fujii:2015jha}. In order to draw conclusions about the
Higgs sector, this precision has to be matched by the Standard Model
prediction. This means that the bottom-quark mass has to be known to
within half a per cent and the charm quark mass to within one per
cent.

Whether this level of precision has already been reached with present
quark mass determinations depends on the way errors are assessed and
propagated when evolving the quark masses from the scale at which they
are determined up to the Higgs boson mass. Following the original
four-loop determination~\cite{Chetyrkin:2009fv,Chetyrkin:2017lif}, the
perturbative uncertainty amounts to $3\,$MeV for the bottom quark and
$2\,$MeV for the charm quark and is therefore negligible. However, this
viewpoint has been challenged
in~\cite{Dehnadi:2011gc,Dehnadi:2015fra}, where the theory
uncertainties are estimated at $10\,$MeV for the bottom quark and
$21\,$MeV for the charm quark. A five-loop quark mass determination
would resolve this disagreement and ensure that future Higgs coupling
measurements are not limited by theory.

\section{Massive QCD at Five Loops}
\label{sec:overview}

The most promising candidates for the first genuine five-loop QCD
calculations are quantities depending on a single scale, which can be
factored out from all Feynman integrals. If this scale is an external
momentum, one arrives at massless propagator-type Feynman
diagrams. For this class of diagrams, a range of dedicated methods
have been developed. For example, integration-by-parts reduction can
be systematised for specific diagram topologies exploiting the
triangle~\cite{Chetyrkin:1981qh} and the diamond~\cite{Ruijl:2015aca}
rule. Diagrams can be simplified --- sometimes even reduced to trivial
base cases --- with the help of graphical
functions~\cite{Schnetz:2013hqa}. Thanks to the glue-and-cut
method~\cite{Baikov:2010hf}, all 281 master integrals are
known~\cite{Georgoudis:2021onj}. Still, the fact that there are 64
diagram families with 15 propagators and 20 possible scalar products
poses a considerable combinatorial challenge.

An alternative is to consider problems with a single non-zero internal
quark mass and vanishing external momenta. If all external particles
are massless, there are 34 families of massive five-loop vacuum
diagrams, with 12 propagators and 15 possible scalar products. While
there are only 156 master integrals, most of them remain unknown. In
the following we will focus on this scenario.

\subsection{General Setup}
\label{sec:setup}

Our calculational setup is based on the standard steps of a multiloop
calculation. Diagrams are generated with
\texttt{QGRAF}~\cite{Nogueira:1991ex}. Their families are identified
using custom code~\cite{dynast} based on \texttt{nauty} and
\texttt{Traces}~\cite{mckay2013practical}. We use
\texttt{FORM}~\cite{Vermaseren:2000nd} for inserting the Feynman rules
and simplifying the resulting expressions. The resulting scalar
integrals are reduced to master integrals using integration-by-parts
reduction~\cite{Chetyrkin:1981qh} via Laporta's
algorithm~\cite{Laporta:2001dd}. To this end, we
use~\texttt{crusher}~\cite{crusher} with~\texttt{tinbox}~\cite{tinbox}
for reduction over finite
fields~\cite{Kauers:2008zz,Kant:2013vta,vonManteuffel:2014ixa,Peraro:2016wsq,Klappert:2019emp}.

\subsection{The Quark Condensate}
\label{sec:cond}

The heavy quark condensate $\langle \bar{\psi}\psi \rangle$ appears in
the leading non-analytic contribution of the Operator Product
Expansion~\cite{Wilson:1969zs}, or equivalently, in the asymptotic
small-mass expansion. It has been suggested that its
non-perturbative value can be obtained from a perturbative evaluation
via renormalisation group optimised perturbation
theory~\cite{Kneur:2010ss,Kneur:2011vi,Kneur:2013coa,Kneur:2015dda,Kneur:2020bph}.
Its anomalous dimension is proportional to the
vacuum anomalous dimension $\gamma_0$~\cite{Spiridonov:1988md}, viz.
\begin{equation}
  \mu^2 \frac{d}{d\mu^2} m\langle \bar{\psi}\psi\rangle = - 4 m^4 \gamma_0,
\end{equation}
which provides a powerful cross check of the five-loop result for
$\gamma_0$~\cite{Baikov:2018nzi}.

The first two orders of the perturbative expansion of the quark
condensate correspond to the sum of only two vacuum diagrams:
\begin{equation}
  \label{eq:qq_exp}
    \langle \bar{\psi}\psi \rangle =\ \centeredgraphics[width=40pt]{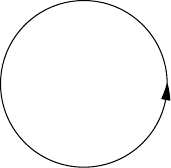}\ +\ \centeredgraphics[width=40pt]{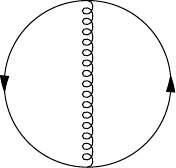}\ +\ \mathcal{O}(\alpha_s^2).
\end{equation}
At five loops, 3451 diagrams contribute. After inserting the Feynman rules, we obtain approximately
400\,000 scalar vacuum integrals with up to 4 dots (i.e.\ propagators raised
by one power) and 4 powers of scalar products in the numerator. We
consider different approaches for the numerical evaluation of the
resulting 156 master integrals.

\subsection{Numerical Evaluation of Master Integrals}
\label{sec:fiesta}

Using \texttt{FIESTA}~\cite{Smirnov:2021rhf} for numerical sector decomposition~\cite{Binoth:2000ps}, we obtain
\begin{equation}
  \begin{split}
    \langle \bar{\psi}\psi \rangle
    \Big|_{\left(\frac{\alpha_s}{\pi}\right)^4} ={} \frac{m^3}{16\pi^2}&\Biggl[\frac{(-3.5 \pm 3.0)\times 10^{{-8}}}{\epsilon^{11}} + \frac{(-2.1 \pm 7.6)\times 10^{{-6}}}{\epsilon^{10}}
    \\
&+ \frac{(0.2 \pm 2.1)\times 10^{{-4}}}{\epsilon^{9}}+ \frac{(-0.5 \pm 7.5)\times 10^{{-2}}}{\epsilon^{8}} + \mathcal{O}(\epsilon^{-7})\Biggr],
  \end{split}
\end{equation}
in $d=4-2\epsilon$ dimensions, concluding that this approach alone is insufficient to obtain a
meaningful result. We observe a loss of approximately two significant
digits per order in the dimensional regulator $\epsilon$, which
suggests that master integrals have to be known with better than
double precision. We have not found any significant improvement using
quasi-Monte Carlo methods~\cite{Borowka:2018goh}, a quasi-finite
basis~\cite{vonManteuffel:2014qoa}, tropical
integration~\cite{Borinsky:2020rqs}, or
\texttt{pySecDec}~\cite{Heinrich:2023til}.

To obtain high-precision results for the master integrals, we use two
approaches. In the first approach, we raise one propagator power in
the master integrals to a symbolic power $x$ and obtain a coupled set
of difference equations via integration-by-parts reduction. This
system is solved numerically through a truncated factorial series
ansatz inserting recursively determined boundary conditions for
$x\to \infty$~\cite{Laporta:2001dd,Luthe:2016sya}, where the integrals
reduce to lower loops.

Alternatively, we evaluate the master integrals via numeric
integration over a loop momentum, where the integrand is a
propagator-type integral. It is then straightforward to evaluate the
angular integral. The momentum routing can always be chosen such that
the loop momentum flows through a massive line. Since all fermion
lines are closed, this means that the propagator-type integrand has no
massless cuts and is therefore infrared finite. Defining the loop integral measure as $[dq] \equiv  \frac{d^dq}{i \pi^{d/2}}$, one arrives at the symbolic form
\begin{equation}
  \label{eq:master}
  T = -\int_0^{-\infty} \frac{dq^2}{\Gamma(2-\epsilon)} \frac{(-q^2)^{1-\epsilon}}{(m^2-q^2)^a} P_0(q^2)\,,
\end{equation}
where $T$ denotes the original vacuum integral, $a$ the power of the
propagator with momentum $q$, and $P_0$ the remaining propagator-type
integral after removing said propagator. The integrand can be made
ultraviolet finite by either introducing a suitable subtraction term
or by choosing $a$ sufficiently large. In the latter case, the
corresponding master integral (where typically $a=1$) can be computed
from $T$ via integration-by-parts reduction.

To compute the integrand $P_0$ we derive a set of differential
equations for the propagator-type master integrals
$P_i$~\cite{Kotikov:1990kg,Remiddi:1997ny}, which we solve for
$q^2 \to 0$ and $q^2 \to -\infty$ with generalised power series
ansätze
\begin{align}
  \label{eq:exp_le}
  P_i ={}& \sum_{k=0}^{N-1} c_{ik} q^{2k} + \mathcal{O}(q^{2N}), \\
  \label{eq:exp_he}
  P_i ={}& \sum_{n=0}^{\# \text{loops}} \sum_{k=k_0}^{N-1} d_{ikn} \left(-\frac{1}{q^2}\right)^{k+n\epsilon} + \mathcal{O}\left(\frac{1}{(-q^2)^N}\right).
\end{align}
The boundary conditions $b_{i,0}$ and $b_{i,k_0,n}$ correspond to
products of known massive vacuum
diagrams~\cite{Schroder:2005va,Chetyrkin:2006bj} and massless
propagators~\cite{Baikov:2010hf,Lee:2011jt} with at most four
loops. After subtracting the logarithmic high-energy contribution and
performing a conformal mapping
$q^2 \to \frac{4\omega}{(1+\omega)^2} m^2$ we construct a
high-precision Padé approximation from the expansion
coefficients~\cite{Baikov:1995ui,Baikov:2013ula,Maier:2017ypu}. A
similar procedure was originally proposed in~\cite{Faisst:2004kz}.

\begin{figure}[htb]
  \centering
    \includegraphics[width=0.45\linewidth]{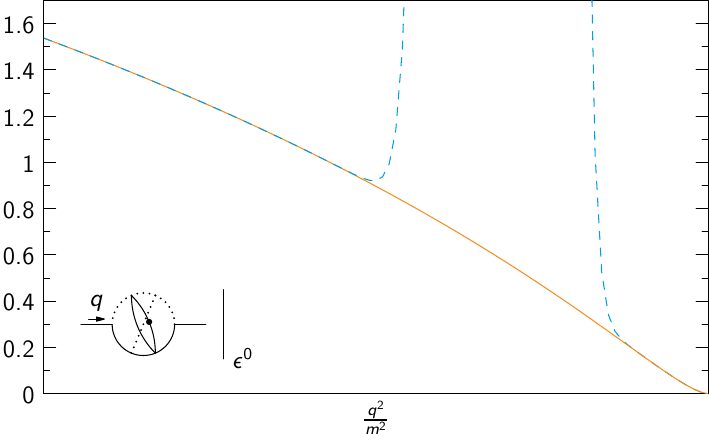} \includegraphics[width=0.45\linewidth]{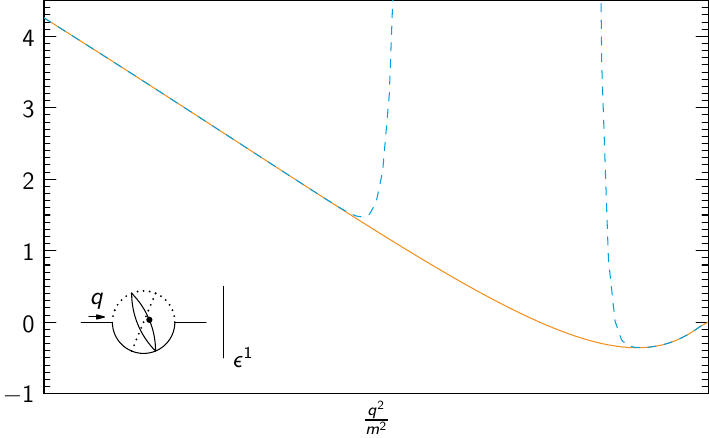}
    \caption{%
      Padé approximations for the leading two coefficients in
      an expansion around $d=4$ dimensions for a four-loop
      propagator-type diagram.
    }
    \label{fig:pade}
\end{figure}

As an example, we consider the following vacuum diagram, routing the
numerical integration momentum through the bottom-most line:
\begin{equation}
    \centeredgraphics[width=40pt]{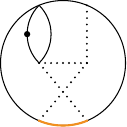} = -\int\limits_0^{-\infty} \frac{dq^2}{\Gamma(2-\epsilon)}\ (-q^2)^{1-\epsilon}\quad \centeredgraphics[width=40pt]{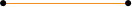} \quad \times \quad \centeredgraphics[width=80pt]{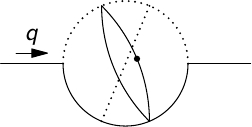}
  \end{equation}
  We derive the differential equations for the four-loop propagator in
the integrand and insert the ansätze in equations~\eqref{eq:exp_le},
\eqref{eq:exp_he}. For $N=20$ we obtain the Padé approximations shown
in figure~\ref{fig:pade}.

Numerical integration then yields
\begin{equation}
 \centeredgraphics[width=20pt]{T5L10lF12}\times e^{5\gamma_E \epsilon} ={} 5.8125309358416596949 - 31.572349480122869826\epsilon + \mathcal{O}(\epsilon^2).
\end{equation}
Changing the integration contour
in equation~\eqref{eq:master} to the line from $0$ to $-i\infty$ changes the
result by less than $10^{-19} + 2\times 10^{-18}\epsilon$. Using
$N=25$ expansion terms reproduces the result within this
uncertainty. We also find good agreement with a numerical evaluation
using \texttt{FIESTA}, which yields $5.81409 - 31.58070(87) \epsilon$.

\subsection{Decoupling}
\label{sec:decoupling}

The presence of heavy quarks in virtual corrections is problematic in
perturbative QCD in calculations where all other energy scales are
much smaller than the heavy quark mass. On the purely practical level,
diagrams with massive internal lines are notoriously hard to
calculate. What is more, potentially large logarithms
$\ln \frac{E}{m_Q}$, where $m_Q$ is the heavy quark mass and $E$ a
typical energy for the process, can spoil the perturbative convergence
in the $\overline{\text{MS}}$ scheme. The solution is to integrate out
the heavy quark, i.e.\ to construct an effective $n_l$ flavour theory,
where $n_f = n_l + 1$ is the original number of quark flavours. The
couplings can be related via
\begin{equation}
  \label{as_dec}
  \alpha_s^{(n_l)} = \alpha_s^{(n_f)} \frac{\Delta_{cgg}}{\Pi_c^2 \Pi_g},
\end{equation}
where $\Pi_c$ and $\Pi_g$ are the (scalar) ghost and gluon polarisation
functions. $\Delta_{cgg}$ is the quantum correction to the truncated
1PI ghost-gluon vertex $\Gamma_{cgg}$, viz.
\begin{equation}
\Gamma_{cgg} = \Gamma^{(0)}_{cgg}(1 + \Delta_{cgg}),
\end{equation}
where $\Gamma^{(0)}_{cgg}$ is the tree-level vertex. All Green
functions are evaluated for vanishing external momentum and vanishing
light quark masses. This decoupling relation is known to four-loop
order~\cite{Schroder:2005hy}.

At five-loop order, we obtain 131\,860\,803 scalar vacuum integrals
with up to 7 dots and 6 scalar products. The reduction is currently
ongoing, with 31 out of 34 integral families completed.

\subsection{Heavy Quark Masses}
\label{sec:mQ}

The currently most precise determinations of heavy quark masses are
based on sum rules. Let us define (inverse) moments $\mathcal{M}_n$ of the ratio
$R_Q(s) = \frac{\sigma(e^+e^- \to Q \bar{Q})}{\sigma(e^+e^- \to \mu^+
  \mu^-)}$ as
\begin{equation}
  \mathcal{M}_n = \int^\infty_{s_0} ds\,\frac{R_Q(s)}{s^{n+1}} = \frac{12\pi^2}{n!} \left[\left(\frac{d}{dq^2}\right)^n \Pi_Q(q^2)\right]_{q^2=0},
\end{equation}
where the proportionality to a derivative of the heavy-quark
contribution $\Pi_Q$ to the vacuum polarisation follows from
a dispersion relation. These derivatives at vanishing external
momentum can be evaluated perturbatively in terms of massive vacuum
diagrams. The quark mass can then be extracted by comparing the
calculated moments to moments obtained from the experimentally
measured $R_Q$ ratio or moments simulated on the lattice.

\begin{figure}[htb]
  \centering
\begin{tabular}{cccccc}
  \includegraphics[width=50pt]{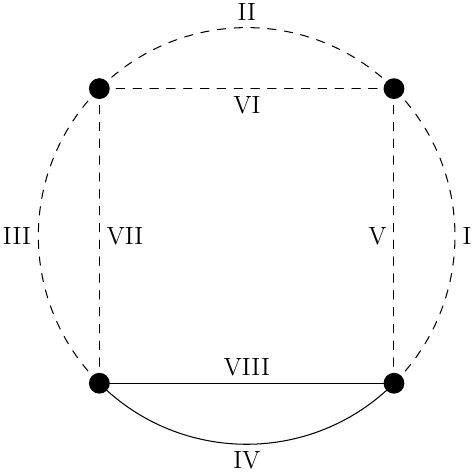} & \includegraphics[width=50pt]{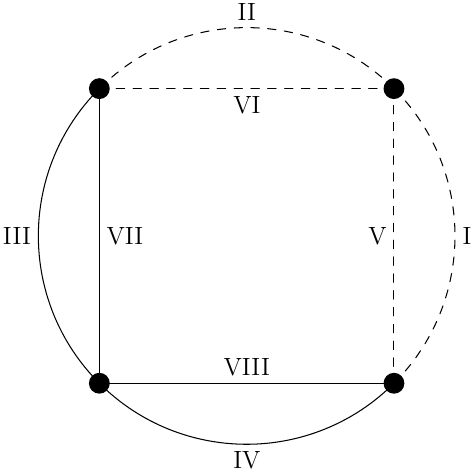} & \includegraphics[width=50pt]{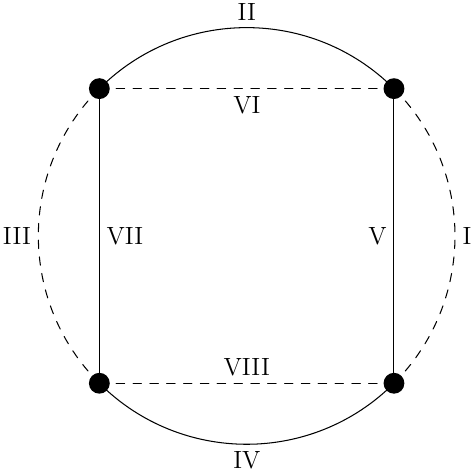} & \includegraphics[width=50pt]{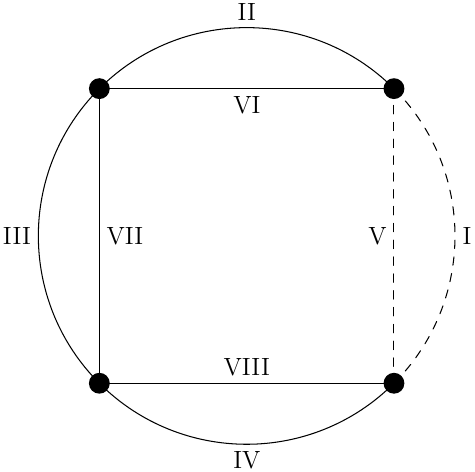} & \includegraphics[width=50pt]{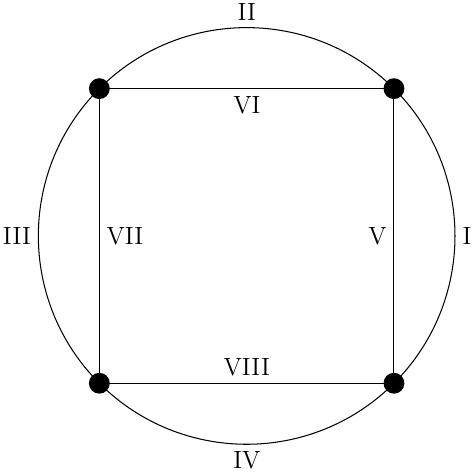} & \includegraphics[width=50pt]{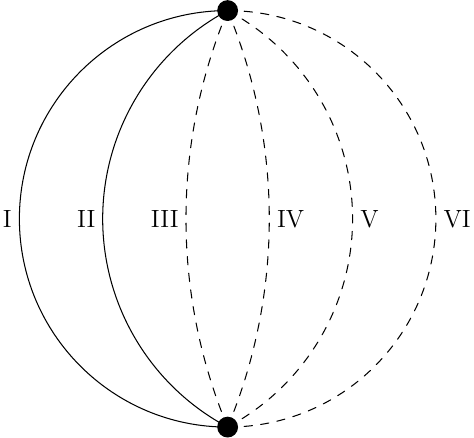} \\
  \includegraphics[width=50pt]{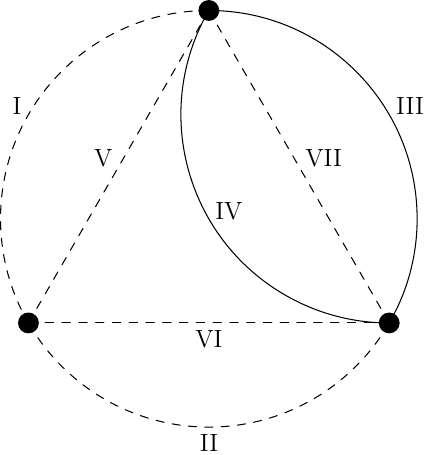} & \includegraphics[width=50pt]{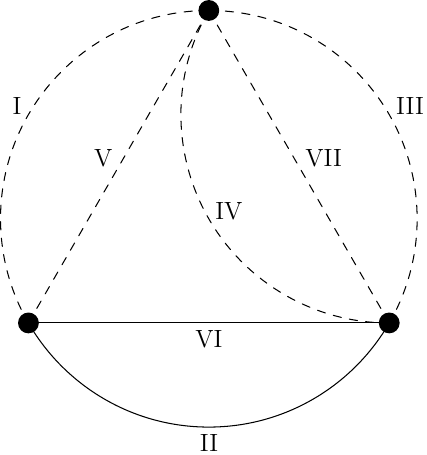} &\includegraphics[width=50pt]{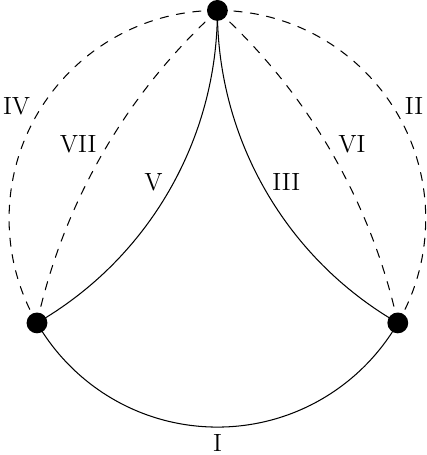} & \includegraphics[width=50pt]{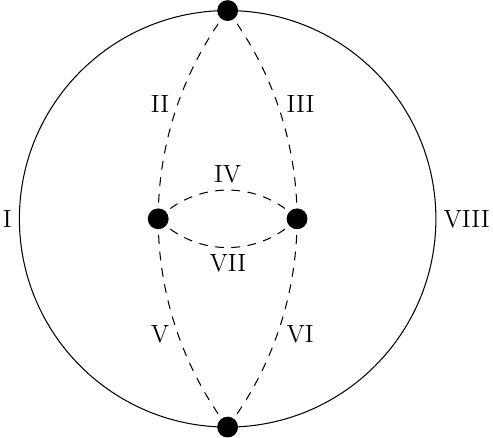} & \includegraphics[width=50pt]{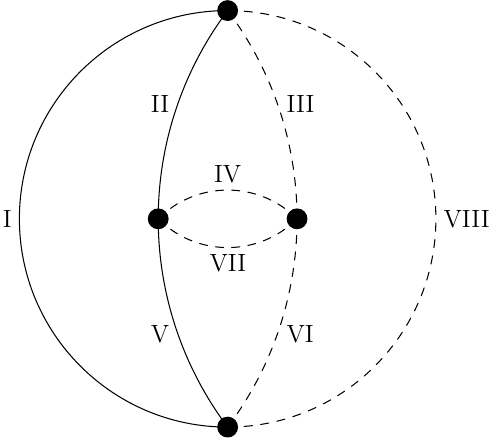} & \includegraphics[width=50pt]{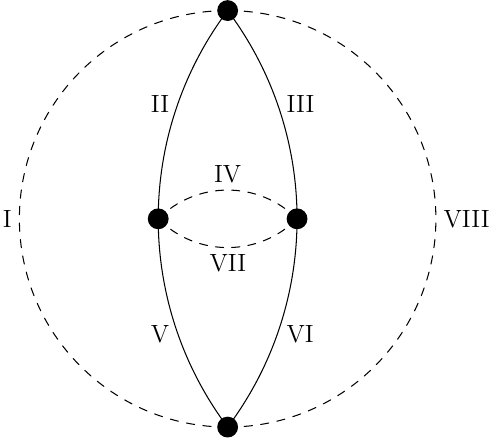}
\end{tabular}
  \caption{Master integrals originating from vacuum diagrams with at least two massless or at least three massive quark loops. }
  \label{fig:masters_nfx}
\end{figure}

We have evaluated a gauge-independent subset of the five-loop
contribution to the first moment. Concretely, for $n_h$ degenerate
massive quarks and $n_l$ massless quarks, we have determined the
contributions proportional to $n_h^4, n_h^3 n_l$, as well as the
contribution from all diagrams with two or three massless quark
loops. Using the setup outlined in section~\ref{sec:setup} we arrive
at a result expressed in terms of the master integrals depicted in
figure~\ref{fig:masters_nfx}.

Evaluating the master integrals with sector decomposition yields
  \begin{equation}
    \begin{split}
    {\cal M}_1^{5\text{ loop}} ={} \frac{3\pi^2}{m_Q^2} \left(\frac{\alpha_s}{\pi}\right)^4 n_h C_F \Big[
      & 0.6\,T_F^3 n_l^3 + 1.2\,T_F^3 n_l^2 n_h + 0.9\,T_F^3 n_l n_h^2\\
      &+ 0.2\,T_F^3 n_h^3 + (C_F - 5 C_A)T_F^2 n_l^2 + \dots
    \Big].
  \end{split}
\end{equation}

\section{Conclusions}
\label{conclusions}

While there is no complete genuine five-loop QCD result so far, a
number of calculations are well underway. There is steady progress
towards five-loop determinations of the heavy quark condensate, the
decoupling coefficients, and the masses of charm and bottom quarks from
sum rules.

The biggest remaining challenge seems to be a numerical high-precision
evaluation of the master integrals corresponding to massive vacuum
diagrams. Here, we use a combination of sector decomposition,
recurrence equations, and direct integration over one loop momentum.

\section*{Acknowledgements}
\label{sec:ack}

Y.S. acknowledges support from ANID under FONDECYT project No. 1231056.

\bibliographystyle{JHEP}
\bibliography{biblio.bib}

\end{document}